\documentclass[%
 reprint,%
 amssymb, amsmath,%
 aip,apl,%
]{revtex4-1}
\usepackage{graphics}

\usepackage{graphicx}
\usepackage{epsfig}
\usepackage{docs}%
\usepackage{bm}%
\usepackage[colorlinks=true,linkcolor=blue]{hyperref}%
\expandafter\ifx\csname package@font\endcsname\relax\else
 \expandafter\expandafter
 \expandafter\usepackage
 \expandafter\expandafter
 \expandafter{\csname package@font\endcsname}%
\fi
\begin{document}

\title{Polyaniline ($C_3N$) nanoribbons: Magnetic metal, Semiconductor, and Half-Metal}%

\author{Meysam Bagheri Tagani}%
\email{m${\_}$bagheri@guilan.ac.ir, Corresponding author}
\author{Sahar Izadi Vishkayi}
\affiliation{Department of physics, Computational Nanophysics Laboratory (CNL), University of Guilan, P.O.Box 41335-1914, Rasht, Iran}%

\date{May 2018}%

\begin{abstract}

Two-dimensional polyaniline sheet has been recently synthesized and found that it is a semiconductor with indirect band gap. Polyaniline nanoribbons decomposed from two-dimensional polyaniline sheet ($C_3N$ sheet) are investigated using density functional theory. The existence of nitrogen atoms in the edge of the ribbons increases stability and magnetization of the ribbons and make them different from graphene nanoribbons. Unsaturated nanoribbons are magnetic metals so that armchair $C_3N$ nanoribbons are gap-less spin semiconductors in the antiferromagnetic state and half-metals in the ferromagnetic state. A transition from metal to semiconductor is observed in the armchair $C_3N$ nanoribbons when the edge atoms are passivated by hydrogen. The band gap of hydrogen saturated armchair $C_3N$ nanoribbons can be controlled using an external transverse electric field so that its magnitude is dependent on the direction of the electric field. Being metal or semiconductor in hydrogen saturated zigzag $C_3N$ nanoribbons is strongly dependent on the edge atoms so that just ribbons having nitrogen atoms in both edges are semiconductor. An external electric field cannot induce any spin polarization in the zigzag nanoribbons which is in contrast with what was observed in zigzag graphene nanoribbons.

\end{abstract}

\maketitle


\section{Introduction}

\par Discovery of graphene as the first two-dimensional (2D) material created a great revolution in industry and science \cite{graphene}. Unique properties of graphene like massless Dirac fermions\cite{mdf}, high thermal and electric conductivity \cite{tcg1, tcg2, ecg1}, and quantum Hall effect \cite{qhf1, qhf2, qhf3} made it most popular material in condensed matter in the last decade. After its discovery, people started to ask about synthesizing other 2D materials.  Answer to these questions led to successful synthesis of silicene \cite{silicene}, germanene \cite{germanene}, stanene\cite{stanene}, black phosphorene \cite{bp}, and borophene \cite{b1, b2} which are single element two-dimensional materials. Graphene, silicene, and germanene have nearly same properties, however, increase of atomic radius induces some buckling in silicene and germanene sheets. Black phosphorene is a semiconductor which makes it distinct in mentioned 2D family. Borophene, a monolayer of boron atoms, has significant differences with other 2D materials. Experimental and theoretical investigations have shown that different phases of borophene can be grown on suitable substrate \cite{b2,b3,b4}.

\par Very recently, 2D polyaniline has been added to two-dimensional family \cite{Mahmood} consisting of six carbon atoms and two nitrogen atoms in a hexagonal lattice with empirical formula of $C_3N$ as shown in Fig.1a. The structure has a Dirac point below the Fermi level and its electric conductivity is $0.72 S/cm$. Next investigations showed that $C_3N$ sheet is an indirect semiconductor\cite{Makaremi}. Liu et.al synthesized a polyaniline crystal with thickness of $0.8 nm$ \cite{Liu}. Thermal conductivity of $C_3N$ sheet is less than graphene making it a potential candidate for thermoelectric applications \cite{ Kumar, Gao}. Li and coworkers showed that a transition from semiconductor to metal is happened when the thickness of $C_3N$ sheet is increased from one layer to three ones \cite{Li}. Although there is superficial similarity between graphene and $C_3N$ sheet, existence of nitrogen atoms obtains outstanding differences in electronic properties. Electrical and mechanical properties of $C_3N$ nanostructures need more investigations in the future.

 \par Convert of 2D sheets to nanoribbons produces fundament changes in electronic and transport properties of them. Therefore, nanoribbons as one-dimensional materials are very important. Cutting direction, nanoribbon width, functionalizing of ribbons' edge significantly affect transport properties of the nanoribbons. There are two typical graphene nanoribbons (GNR) as armchair graphene nanoribbons (AGNRs) and zigzag graphene nanoribbons (ZGNRs) \cite{GNR1, GNR2}.
 Band gap oscillation with ribbons' width is observed in AGNRs. On the other hand, ground state of hydrogen saturated ZGNRs is antiferromagnetic (AFM) and they are semiconductor. It was shown that ZGNRs can be changed to half-metals using an external transverse electric field \cite{EG1, EG2}.  In addition of  quantum confinement effect coming from finite width of the ribbon, existence of two different atoms in the $C_3N$ structure leads to interesting phenomena which are absent in GNRs. In this research, we study electric and magnetic properties of $C_3N$ nanoribbons using density functional theory (DFT). Not only the effect of ribbon width and edge profile is studied but also influence of edge passivation is investigated in details. We find that unsaturated armchair $C_3N$ nanoribbons can be magnetic or half-metal dependent on its width and edge atoms. Armchair ribbons having nitrogen atoms in both edges are gap-less semiconductors in AFM configuration, whereas they become half-metal in ferromagnetic (FM) configuration. Our analysis shows that the gap-less semiconductors can be converted to half-metal using suitable external transverse electric field. A transition from magnetic metal to semiconductor is observed when armchair ribbons are terminated by hydrogen. These observations are completely different from what we knew about AGNRs and emphasis potential of $C_3N$ nanoribbons for electronics and spintronics applications. Zigzag $C_3N$ nanoribbons are magnetic metals when they are not passivated by hydrogen atoms. A very interesting point about hydrogen saturated $C_3N$ nanoribbons is a transition from metal to semiconductor when both edges have nitrogen atoms. This is in contrast with hydrogen terminated ZGNRs, whose ground state is AFM. Computational method is presented in next section. Section 3 is devoted to results. Effect of edge configuration, ribbon's width, and edge passivation is discussed in details. And some sentences are given as a summary at the end of article.

\section{Simulation method and formalism}
 \label{Model}
 \par All calculations were performed using SIESTA package based on density functional theory \cite{siesta}. The cut-off energy was set to be $100 Ha$. A unit cell of $C_3N$ sheet was sampled using $(61 \times 61 \times 1)$ Monkhorst-Pack mesh \cite{Monkhorst}, while  $100$ K-points were used for sampling of first Brillouin zone of a ribbon. General gradient approximation (GGA) with Perdew-Burke-Ernzerhof exchange-correlation functional (PBE) \cite{PBE} and norm-conserving pseudopotential were employed to describe core electrons. A $30 \AA$ vaccume layer was chosen to neglect the interaction of the ribbon with its image. Double-zeta-single polarized basis set (DZP) was used. Thirteen orbitals were employed for each carbon and nitrogen atom consisting of two sets of orbitals for $s$  type, two sets for $p$ type, and one set for $d$ type with cut-off radius of $2.4 \AA$, $2.9 \AA$, and $2.9 \AA$ for carbon atoms and $2.144 \AA$, $2.624 \AA$, and $2.624 \AA$ for nitrogen atoms, respectively. All ribbons were fully optimized with a force tolerance of $0.001 eV/ \AA$. For all ribbons, probability of spin polarization at the edges of a ribbon was considered.
\par A transverse electric field $E_{external}={V_{external}/{d}}$ is applied across the ribbons where $d$ is the lattice constant across the ribbon. Poisson equation was solved using dirichlet boundary condition across the ribbon and periodic boundary condition along the ribbon. Spin population of each atom was calculated using mulliken population analysis \cite{mul}.
\begin{figure}[htb]
\begin{center}
\includegraphics[height=80mm,width=90mm,angle=0]{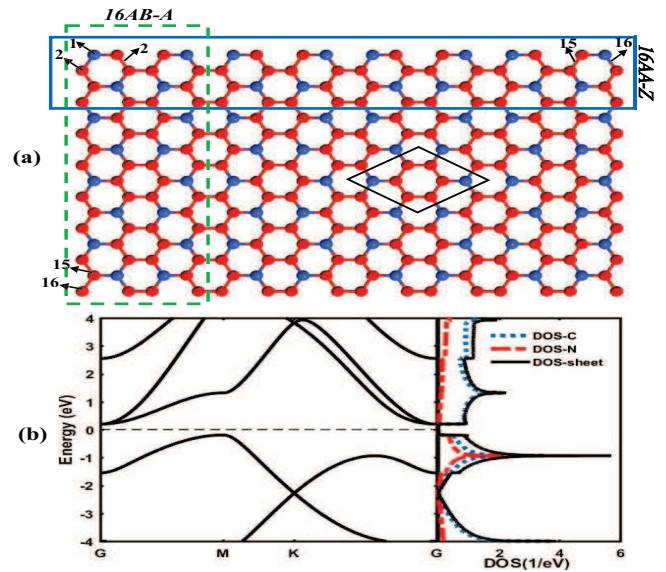}
\caption{\label{fig:1}  (a) 2D $C_3N$ sheet: unit cell of the sheet is shown by black rhombus. Unit cell of 16AA-Z ribbon is shown by solid rectangular and unit cell of 16AB-A ribbon is shown by dashed rectangular. Red and blue balls are carbon and nitrogen atoms, respectively. (b) Band structure, left panel, and DOS, right panel of the sheet. Carbon and nitrogen contribution in the DOS is also plotted. }
\end{center}
\end{figure}
\begin{figure}[htb]
\begin{center}
\includegraphics[height=80mm,width=85mm,angle=0]{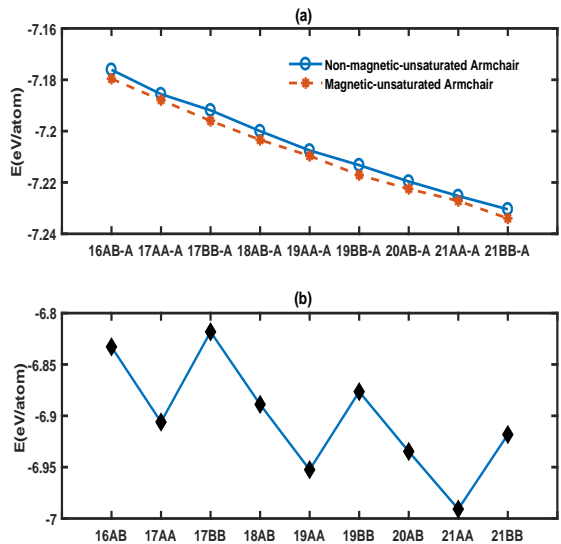}
\caption{\label{fig:2} (a) The formation energy of \emph{NAB-AR}s for magnetic and non-magnetic configurations. (b) The formation energy of \emph{NAB-AHR}s which are non-magnetic. }
\end{center}
\end{figure}
\section{Simulation results}
\label{Numerical results}
\par Figure 1a shows the structure of $C_3N$ sheet with a hexagonal lattice. Each unit cell is composed of six carbon and two nitrogen atoms. Lattice constant is equal to $4.87 {{\AA}}$ which is in good agreement with pervious experimental and theoretical reports \cite{Mahmood,Li}. Band structure and density of states (DOS) of each unit cell is shown in Fig. 1b. The $C_3N$ is a p-type semiconductor with indirect band gap so that maximum of valance band (VBM) is located in $M(1/2,0,0)$ and minimum of conduction band (CBM) is located in $\Gamma$ point. The obtained band gap is $0.4 eV$ which is consistent with previous calculation based on GGA approximation and experimental investigations\cite{Makaremi,new}. The $C_3N$ sheet can be considered as a p-type semiconductor because the VBM is located near the Fermi level.  The von Hove singularity is observed in the DOS located $-1 eV$ below the Fermi level. The energy band gap is also observed in the DOS as zero around the Fermi level. A Dirac cone is observed at $-2.5 eV$ below the Fermi level which is supported by a saddle zero in the DOS plot. Our investigations show that the density of the states located below the Fermi level are created by equal contributions of carbon and nitrogen atoms. In contrast, just carbon atoms participate in the density of states slightly above the Fermi level. The existence of Dirac cone, band gap, von Hove singularity, and anisotropic contributions of carbon and nitrogen atoms in the DOS make $C_3N$ nanostructures very interesting which will attract a lot attention in the future. The similarity between our results and others confirms that the simulation details employed in this research have enough accuracy.
\begin{figure}[htb]
\begin{center}
\includegraphics[height=90mm,width=85mm,angle=0]{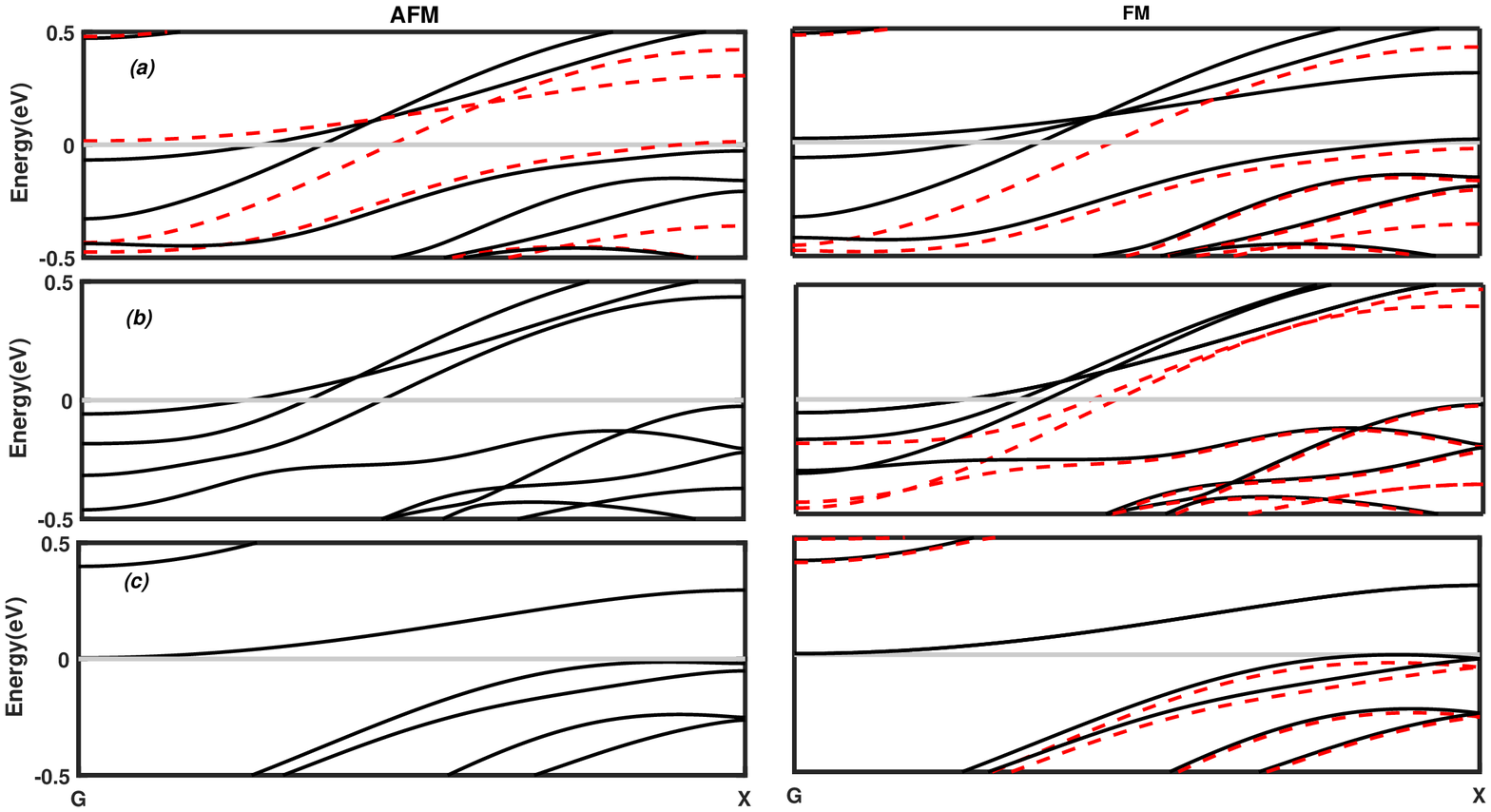}
\caption{\label{fig:3} The band structure of (a) \emph{18AB-AR}, (b) \emph{19AA-AR}, and (c) \emph{19BB-AR} in FM and AFM configurations. Spin-up is denoted by dashed line and spin-down is shown by solid line. Gray line shows the Fermi energy.}
\end{center}
\end{figure}

\begin{figure}[h]
\begin{center}
\includegraphics[height=60mm,width=60mm,angle=0]{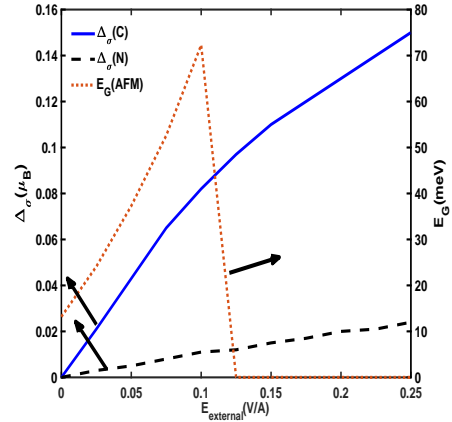}
\caption{\label{fig:4} The net magnetization of edge atoms for carbon (solid line) and nitrogen (dashed line) versus external electric field in AFM \emph{17BB-AR}. The variation of band gap (dotted line) versus electric field is also drawn. }
\end{center}
\end{figure}
\par Unlike graphene, cutting a $C_3N$ monolayer along x or y-direction leads to two different edge configurations: edge with just C atoms (like graphene) or edge with equal number of C and N atoms. We show the former with $A$ and the latter with $B$, so a unit cell of an armchair $C_3N$ nanoribbon is shown by \emph{NUV-AR} where $U (V) =A$  or $B$ and $N$ stands for the number of atoms in a row across the ribbon, see Fig. 1a. There are two different edge configurations for even $N$ i.e. $AA$ and $BB$ and just one configuration for odd $N$ i.e. $AB$. It is clear from Fig. 1a that a unit cell of $C_3N$ nanoribbon is wider than graphene one. We analyze the $C_3N$ nanoribbons with two different scenarios: unsaturated ribbons and ribbons saturated by hydrogen atoms. For unsaturated edges, edge reconstruction is an important issue which can change the final results. So, we also considered a $2\times 1$ supercell for optimization process. The obtained results  do not exhibit a significant change in comparison with results prepared from optimization of a unit cell. In addition, we made an Ab-initio molecular dynamic simulation at room temperature to analyze edge reconstruction with more details.  First, we analyze unsaturated armchair $C_3N$ nanoribbons which can be synthesized in ultra-high vacuum condition. Figure 2 shows formation energy of the ribbon, $E_{f}=\frac{E_{ribbon}-N_{N}E_{N}-N_{C}E_{C}}{N_{N}+N_{C}}$ where $E_{ribbon}$, and $E_{N}$ ($E_{C}$) are the total energy of a unit cell of the ribbon, and energy of an isolated nitrogen (carbon) atom, respectively. The ground state of the unsaturated armchair ribbons is degenerated and magnetic so that ferromagnetic (FM) and antiferromagnetic (AFM) configurations have the same energy. This phenomenon makes armchair $C_3N$ nanoribbons completely different from unsaturated AGNRs because their ground state is non-magnetic. Stability of the ribbons increases with their width which was predictable. Energy difference between the magnetic and non-magnetic ground state of considered \emph{NAA-AR}s is about $160 meV$, while it is $295 meV$ for \emph{NBB-AR}, and as a result, \emph{NBB-AR}s are more stable than \emph{NAA-AR}s. The mentioned energy difference is about $230 meV$ for considered \emph{NAB-AR}s. As a result, the unsaturated armchair ribbons will be magnetic at room temperature. Increase of magnetization in the edges having nitrogen atom comes from an extra electron that is donated to the ribbon from nitrogen atom.
\begin{figure}[h]
\begin{center}
\includegraphics[height=90mm,width=85mm,angle=0]{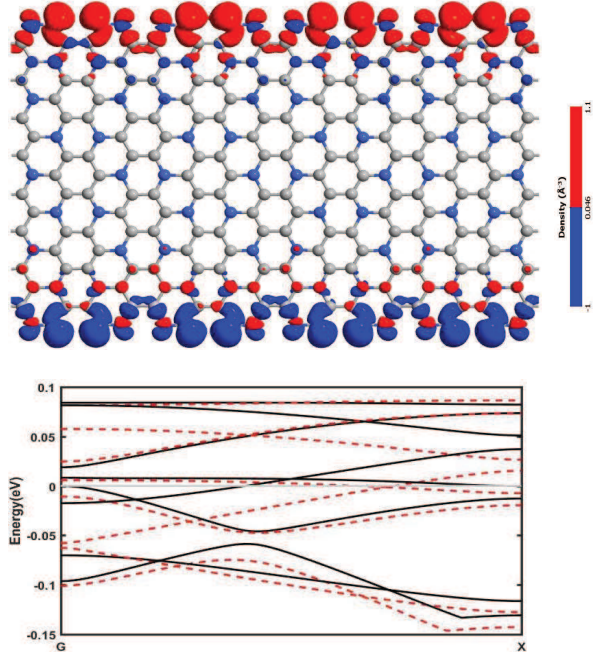}
\caption{\label{fig:5} The spin density (Blue and gray balls are nitrogen and carbon atoms, respectively)
and the band structure (solid (dashed) lines are for spin up (down) bands) of $17BB-AR$ after molecular dynamics simulation for $10 ps$.}
\end{center}
\end{figure}
\par Figure 3 shows band structure of \emph{18AB-AR}, \emph{19AA-AR}, and \emph{19BB-AR} in FM and AFM configuration. A very interesting observation for \emph{18AB-AR}  is spin band splitting in AFM state which was not observed in AGNRs. The spin splitting is attributed to asymmetry in two edges of the ribbon. Indeed, spatial anisotropy in the edges gives rise to the breaking spin degeneracy in energy space. Recently, we reported a similar effect in $\beta_{12}$ borophene nanoribbons \cite{Sahar}. From mulliken population analysis we found that the magnetization of each C (N) atom in the $B$  type edge is equal to $0.7 \mu_{B}$ ($0.15 \mu_B$). The magnetization story is more interesting for $A$ type edge so that the magnetization of each C atom connected to N one is $0.38 \mu_B$ whereas, C atoms which are not connected to N ones are non-magnetic.
 \emph{18AB-AR} is a magnetic metal while, AGNRs are semiconductor. \emph{19AA-AR} is also a magnetic metal in both AFM and FM configurations. There are several bands crossing the Fermi level in AFM configuration so the ribbon cannot convert to a half-metal using an external transverse electric field. In ferromagnetic configuration, the number of bands crossing the Fermi level is more for minority spin carries leading to an asymmetry in transport properties of \emph{NAA-AR}s.
\begin{figure}[h]
\begin{center}
\includegraphics[height=80mm,width=85mm,angle=0]{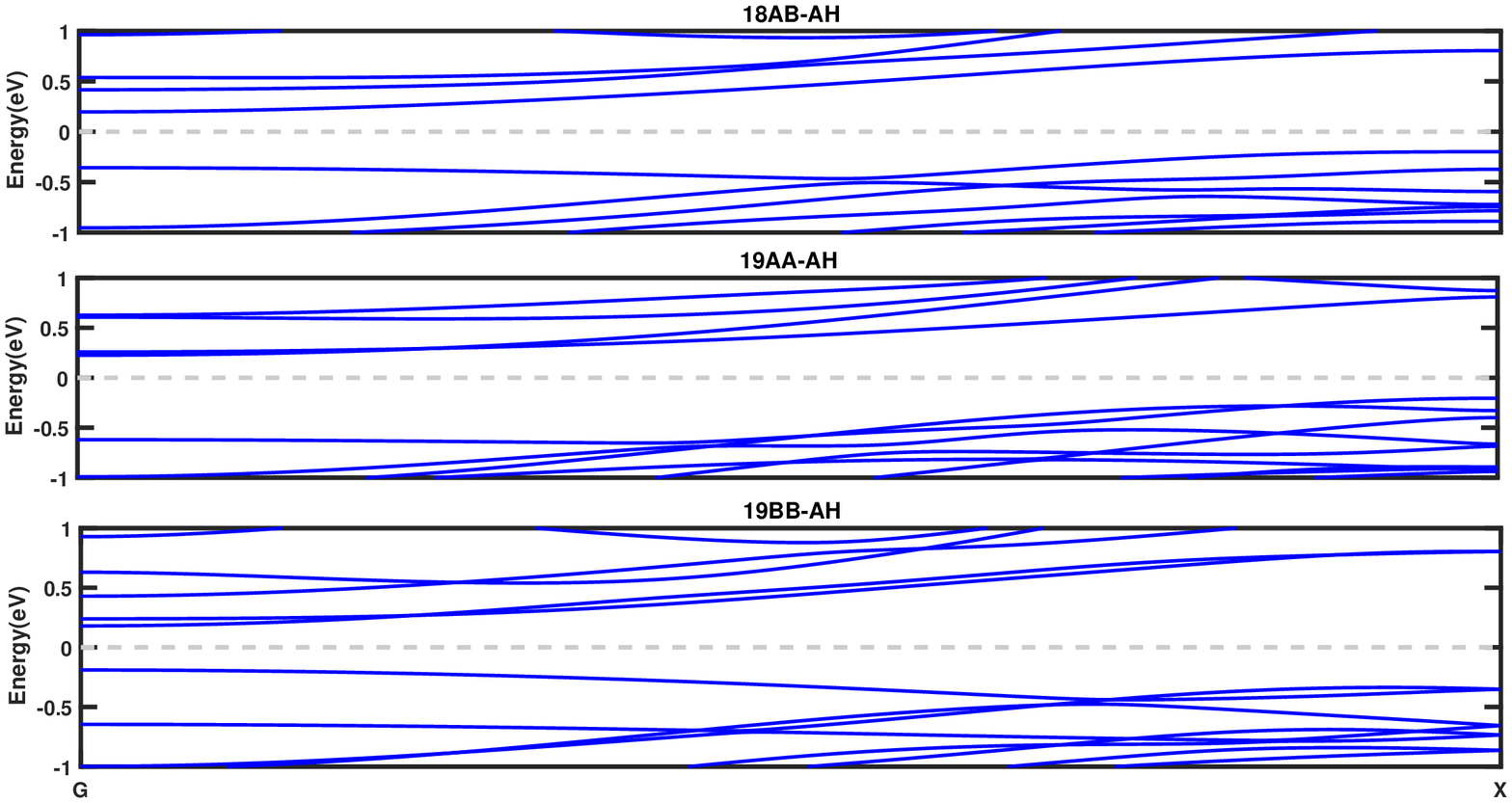}
\caption{\label{fig:6} The band structure of hydrogen passivated armchair nanoribbons. }
\end{center}
\end{figure}
\par The band structure of \emph{(2N+1)BB-AR}s is very interesting and strange. We are faced with a gap-less semiconductor in AFM configuration. Therefore, applying an external transverse electric field can convert the ribbon from semiconductor to a half-metal structure. Note that the zigzag graphene nanoribbons could transform to a half-metal, while here, unsaturated armchair $C_3N$ nanoribbons. Our analysis reveals that the AFM \emph{17BB-AR} changes to a half-metal when external electric field is less than $0.125 V/{\AA}$ and after that the ribbon will be a ferromagnetic metal as shown in Fig.4 . External electric field induces an inhomogeneous spatial distribution of spin in two edges so that a net magnetization, the difference between the magnetization of upper and lower edge atoms, is appeared. Fig.4  shows that the net magnetization increases by the electric field and its effect is more pronounced on carbon atoms. FM \emph{(2N+1)BB-AR}s are intrinsic half-metals with a spin gap equal to $470 meV$. This result makes them a special unit in 2D materials world so that they can be a promising candidate in next-generation spintronic applications. The observed spin gap is a robust feature of the ribbons so that external transverse electric field as high as $E_{external}=0.25V/{\AA}$ makes no change in the gap. The spin density and DOS of the ribbons for higher external electric fields are plotted in supplementary information. It is revealed that applying s stronger electric field shifts the spin band gap above the Fermi level. These results are completely different from what we found in unsaturated AGNRs. Unsaturated AGNRs are semiconductor and external transverse electric field cannot induce any spin polarization in the ribbon. It just shifts the bands so that energy gap is decreased. Band structure of an unsaturated 10AGNR is plotted in Fig. S3 for $E_{external}=0$, and $E_{external}=0.2 V/{\AA}$.
\begin{figure}[h]
\begin{center}
\includegraphics[height=70mm,width=70mm,angle=0]{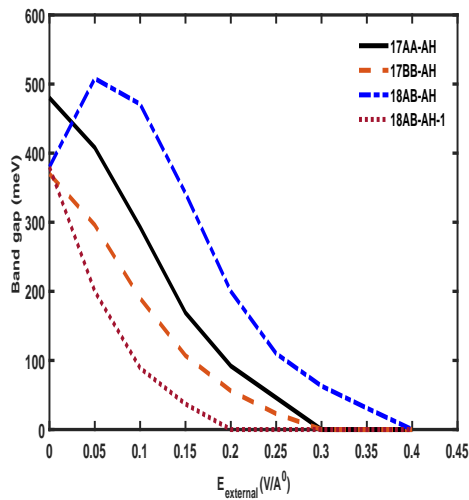}
\caption{\label{fig:7} Variation of band gap versus external electric field. The electric field points from $A$ type edge to $B$ type edge in \emph{18AB-AH} and vice versa in \emph{18AB-AH1}.}
\end{center}
\end{figure}
\par Edge reconstruction is an important issue for unsaturated nanoribbons which can dominate their electronic properties. To have an approximate insight about the effect of finite temperature on the edge configuration of the $(2N+1)BB-ARs$ and their electronic properties, we have used classical molecular dynamics to simulate $17BB-AR$ at $300 K$. A super cell  of $4\times1\times1$ in NVT ensemble is simulated for $10 ps$ with time step of $1 fs$. Final structure shows partial edge reconstruction. However, the obtained structure still prefers to be magnetic. Its spin density is plotted in Fig. 5. As shown in Fig. 5 the ribbon is a magnetic metal in antiferromagnetic state and spin degeneracy is broken due to disordered edge profile.
\par A transition from ferromagnetic metal to semiconductor is observed when the armchair $C_3N$ nanoribbons are passivated by hydrogen, \emph{NUV-AHR}. Fig. 6  shows the band structure of three different classes of the ribbons. \emph{2N-AB-AHR}s are indirect semiconductors so that the VBM is located in $X$ and CBM is in $\Gamma$ point. \emph{(2N+1)AA-AHR}s are also indirect semiconductors, but, the position of valance band maximum is related to the ribbon width. On the contrary, \emph{(2N+1)BB-AHR}s are direct semiconductors so that the gap is located in $\Gamma$ point. The computed band gap of \emph{18AB-AHR} is $380 meV$, while it is $480 meV$, and $370 meV$ for \emph{17AA-AHR}, and \emph{17BB-AHR}, respectively. Results show that the threshold electric field, the external transverse electric field in which the ribbon is transformed from semiconductor to metal, is nearly equal for (\emph{2N+1)AA-AHR} and \emph{(2N+1)BB-AHR}s.  Change of energy gap against external electric field is dependent on the direction of the field in \emph{2NAB-AHR}s which makes them very interesting for field effect applications. Indeed, we found that when the electric field is directed from $B$-type edge toward $A$-type edge, the gap is monotonically reduced. On the other hand, when the direction of electric field is reversed first an increase in the energy gap is observed, then, the energy gap is smoothly reduced toward zero as shown in Fig.7. The observation is a robust feature of \emph{2N-AB-AHR}s family and the threshold electric field and the electric field causing the maximum value of the energy gap are dependent on the ribbon width. Change of energy gap with electric field and band structure of \emph{12AB-AHR} is drawn in Fig. S2. Role of edge passivation in armchair $C_3N$ and graphene nanoribbons is also different. Edge passivation in AGNRs leads to the increase of the band gap and change of the ribbons from indirect semiconductors to direct ones as shown in Fig. S3. In addition, the electric field cannot induce any spin polarization in the AGNRs.
\par In the following, unsaturated (saturated) zigzag $C_3N$ nanoribbons are studied in details. We call them \emph{NUV-Z(H)R} that $N$ denotes the number of dimers across the ribbon as shown in Fig.1a, $U (V)= A, B$ stands for the edge type and $H$ is used for hydrogen terminated ribbons. Ribbons with even $N$ have two different edge configurations i.e. $AA$, and $BB$, while odd $N$ ribbons are $AB$. Formation energy of zigzag ribbons is drawn in Fig.8. First, we analyze unsaturated zigzag ribbons. Results show that all ribbons are magnetic and the ribbons having nitrogen atom in the edge are more stable than \emph{NAA-ZR}s. Interesting point is that ribbons with narrower width and nitrogen atom in the edges are more stable than the wider ribbons with $AA$ configuration. The energy difference between magnetic and non-magnetic state is about $150 meV$ for $AA$, $980 meV$ for $BB$ and $575 meV$ for $AB$ configurations. It means that the unsaturated ribbons will be magnetic at room temperatures. The energy difference obtained for zigzag ribbons shows behavior similar to armchair ones but with more pronounced intensity.  Magnetization of carbon atoms in the edge of $A$ type is $0.3 \mu_B$, whereas it is about $1 \mu_B$ in $B$ type edge. Indeed, existence of nitrogen atom in the edge not only increases stability of the ribbon but also increases the magnetization of the ribbon. Note that the ground state of unsaturated zigzag graphene nanoribbons is antiferromagnetic. In addition, our investigation shows that the magnetization of each edge carbon atom is $1.22 \mu_B$. Existence of nitrogen atom brings significant changes in the electronic properties of zigzag nanoribbons. The formation energy of hydrogen passivated ribbons increases with their width as shown in the inset of Fig. 8. As unsaturated case, ribbons having nitrogen atoms in the edge are more stable than ones with just carbon atoms in the edges.
\begin{figure}[htb]
\begin{center}
\includegraphics[height=90mm,width=90mm,angle=0]{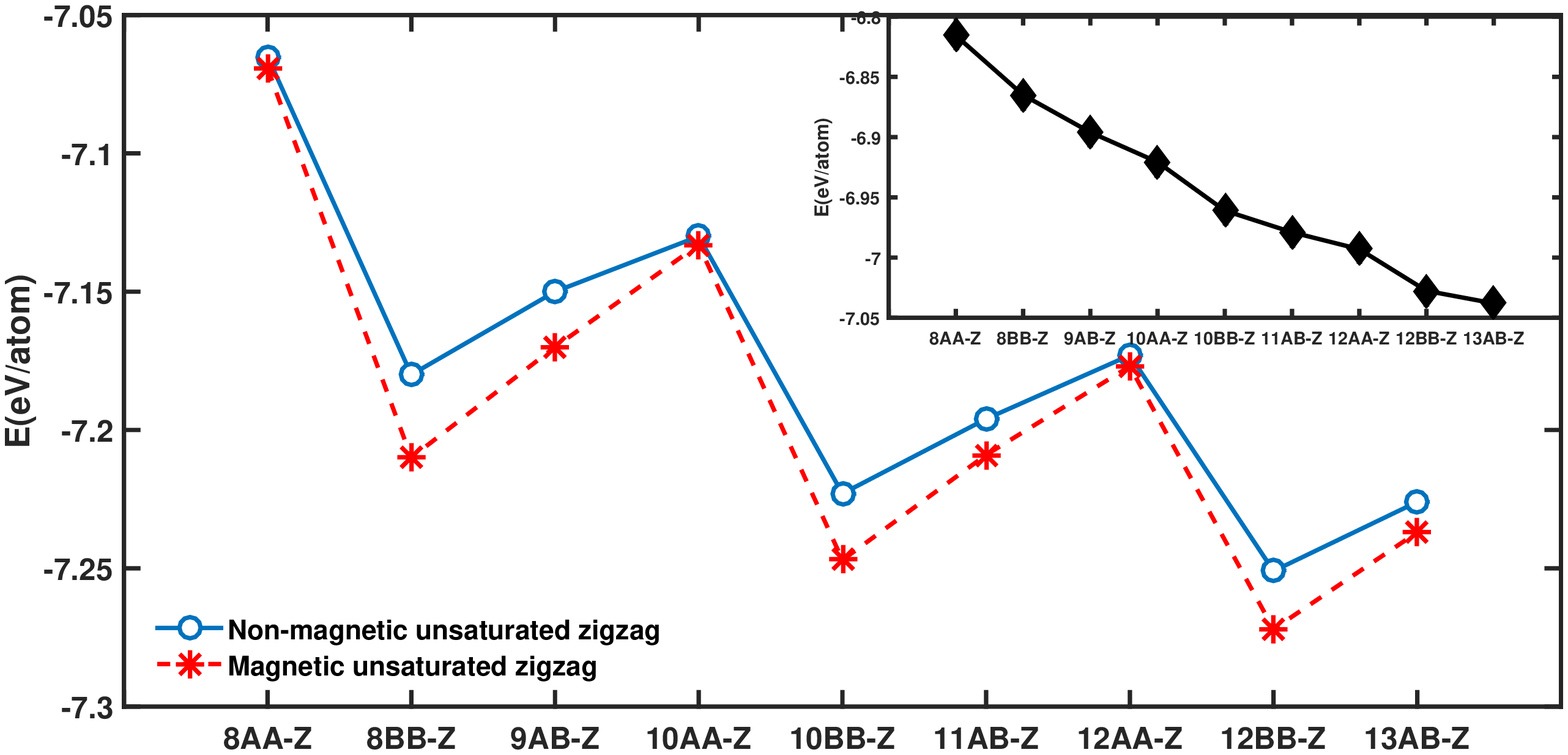}
\caption{\label{fig:8}  The formation energy of \emph{NAB-ZR}s for magnetic and non-magnetic configurations. Inset shows the formation energy of \emph{NAB-ZHR}s which are non-magnetic. }
\end{center}
\end{figure}
\begin{figure}[htb]
\begin{center}
\includegraphics[height=80mm,width=85mm,angle=0]{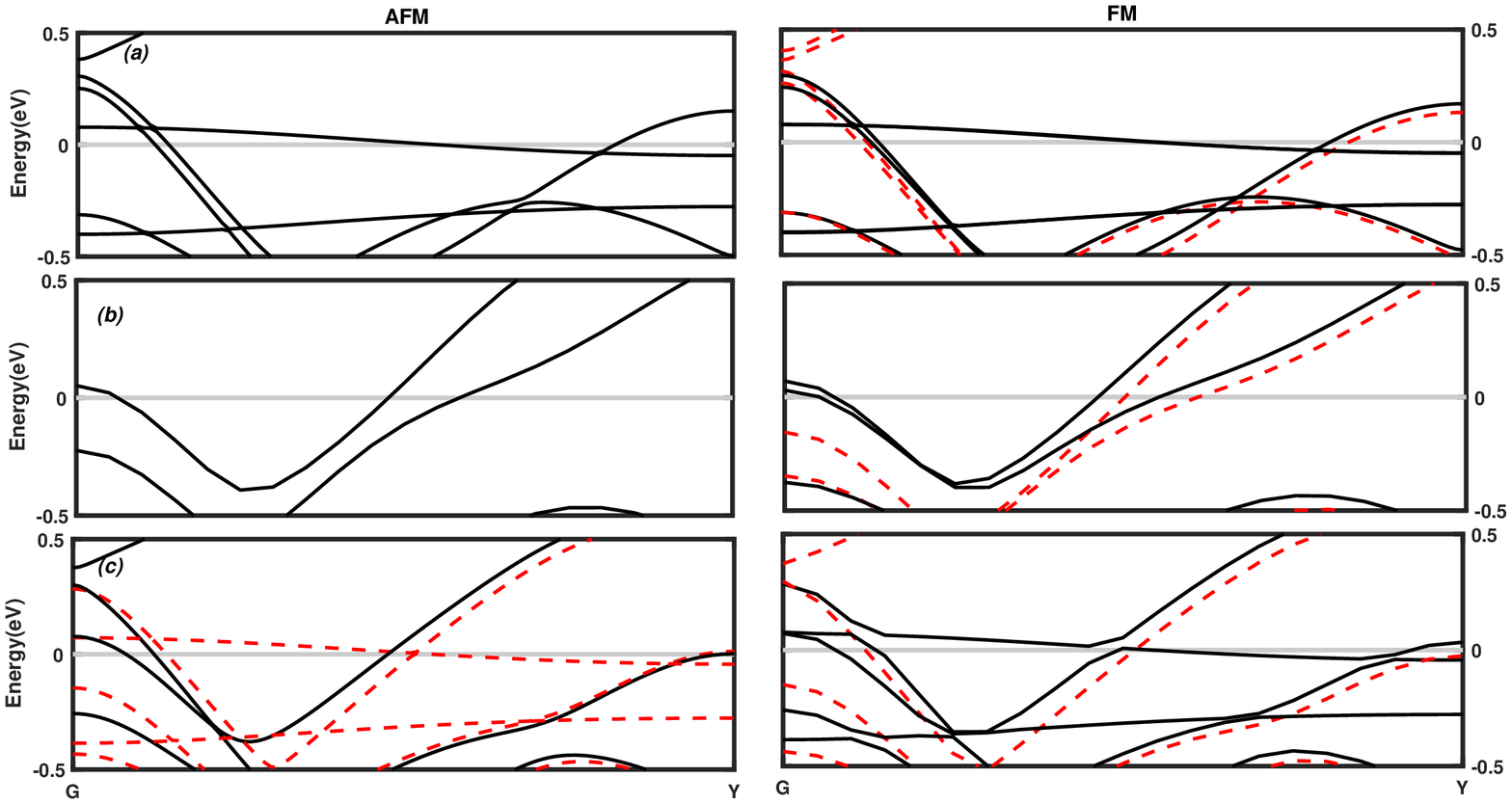}
\caption{\label{fig:9} The band structure of (a) \emph{8AA-ZR}, (b) \emph{8BB-ZR}, and (c) \emph{9AB-ZR} in FM and AFM configurations.
Spin-up is denoted by dashed lines and spin-down is shown by solid lines. Gray lines show the Fermi energy.}
\end{center}
\end{figure}
\par Band structures of three different edge profiles of unsaturated zigzag ribbons are plotted in Fig.9  for both AFM and FM configurations. All unsaturated ribbons are metal, however, the edge profile dominates their electronic properties strongly. There is nearly flat band near Fermi level for $AA$ configurations which comes from minority spins in FM state. There are two bands crossing the Fermi level for each spin component in $BB$ configuration. Unlike armchair ribbons, there is no chance to convert \emph{NBB-ZR}s to a half-metal using a transverse electric field. Anisotropy in the edge profile of the \emph{NAB-ZR}s leads to the breaking of spin degeneracy in the band structure like \emph{NAB-AR}s. The band structure of considered zigzag ribbons shows that the metallic property of the ribbons is strong so that one cannot convert them to a semiconductor or half-metal easily. Comparison of band structures of three different configurations indicates that the flat bands near the Fermi level come from minority spins of A type edge. Unsaturated AFM ZGNRs are direct semiconductors whereas FM ones are ferromagnetic metals as shown in Fig. S3.
\begin{figure}[h]
\begin{center}
\includegraphics[height=85mm,width=85mm,angle=0]{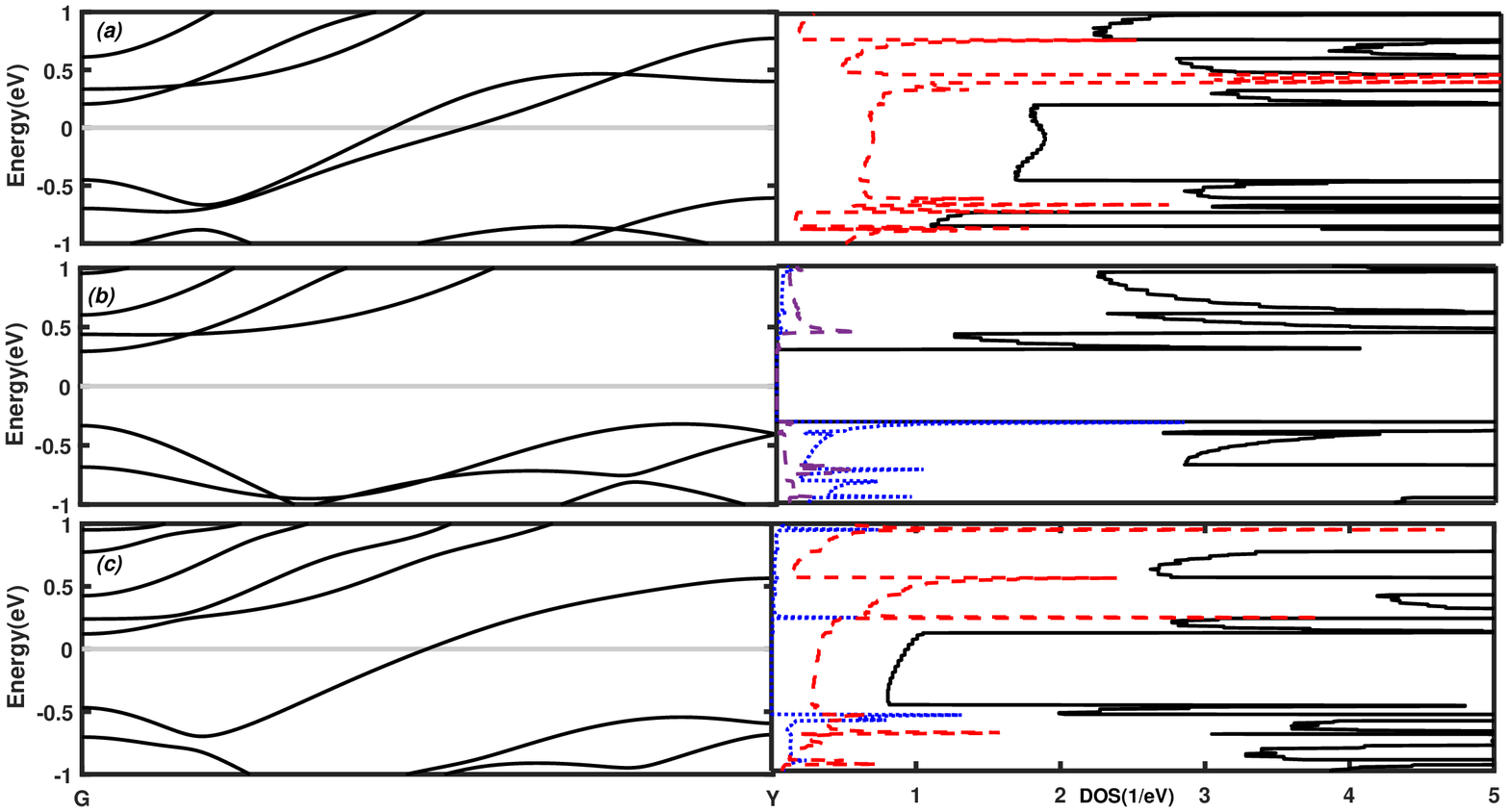}
\caption{\label{fig:10} Left panels: The band structure of (a) \emph{8AA-ZHR}, (b) \emph{8BB-ZHR}, and (c) \emph{9AB-ZHR}. Right panels: Total DOS (solid line), DOS of edge carbon atoms (dashed line), and DOS of edge nitrogen atoms (dotted line).}
\end{center}
\end{figure}
 \par Role of hydrogen passivation for zigzag ribbons is different from armchair ones. We observed that the edge saturation leads to a transition from metal to a semiconductor in armchair ribbons, but here, the behavior is different. Our analysis shows that the zigzag ribbons which have at least one $A$ type edge are still metal after passivation as shown in Fig.10. In return, \emph{NBB-ZHR}s are indirect semiconductors unlike graphene zigzag nanoribbons which are metal. The energy gap is equal to $628 meV$ for $6BB-ZHR$ and $448 meV$ for \emph{14BB-ZHR}. To find the origin of this effect, we calculate the density of states of the edge carbon and nitrogen atoms. Fig. 10 shows that the nitrogen atoms located at the edge of the ribbon open a band gap around the Fermi level, whereas the DOS of carbon atoms is nonzero. Therefore, the ribbons having $A$ type edge exhibit metallic properties and \emph{NBB-ZHR}s are semiconductor. We examine the probability of half-metallicity in semiconductor zigzag nanoribbons in the presence of a transverse external electric field. Spin-polarized calculations show that the passivated ribbons are non-magnetic under external electric field so that there is no chance to convert them to a half-metal. This observation is in contrast with zigzag graphene nanoribbons. The external electric field decreases the energy band gap of the \emph{NBB-ZHR}s as shown in   Fig.11. We fitted the change of the energy band gap versus external electric field for \emph{8BB-ZHR} with a quadratic function as $E_G(eV)= a + bE_{ext}+cE_{ext}^2$ and found $a=616.5 eV$, $b=-362.86 e\AA$, and $c=-3885.71 e\AA^2/V$. Our investigations show that the quadratic dependence of the energy band gap to transverse electric field is independent of the width of the ribbon. Hydrogen terminated ZGNRs are  AFM in the ground state and they are semiconductors with a small gap. Non-magnetic and FM states of hydrogen passivated ZGNRs are metals. In hydrogen terminated zigzag $ C_3N$ nanoribbons ferromagnetic and antiferromagnetic states are degenerated and being metallic or semiconductor is dependent on the existence of nitrogen atoms in the edge of the ribbon. This observation makes $C_3N$ nanoribbons different from GNRs. Unsaturated AFM ZGNRs are direct semiconductors whereas unsaturated FM ZGNRs are ferromagnetic metals as shown in Fig. S3.
\begin{figure}[h]
\begin{center}
\includegraphics[height=80mm,width=80mm,angle=0]{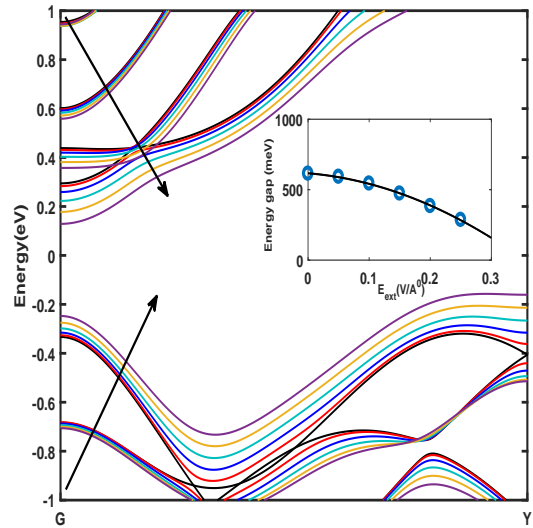}
\caption{\label{fig:11} The band structure of \emph{8BB-ZHR} under an external transverse electric field. Arrows show the increase of electric field strength.  The band gap is reduced by increase the electric field strength. The variation of the band gap versus the electric field is shown in the inset (by circle). The solid line shows  the dependence of the band gap on a quadratic function of the electric field. }
\end{center}
\end{figure}
\section{Conclusion}
\label{conclusion}
We have used density functional theory to study the electronic and magnetic properties of armchair and zigzag Polyaniline ($C_3N$) nanoribbons which are recently synthesized experimentally.
The effect of  edge atoms, ribbon's width, edge passivation, and external transverse electric field was investigated in details. The existence of nitrogen atoms in the structure produces significant differences in comparison with graphene nanoribbons. Bare armchair $C_3N$ nanoribbons can be magnetic metal or half-metal dependent on their edge atoms.The armchair nanoribbons having nitrogen atoms in  both edges are intrinsically  half-metals in the ferromagnetic state and can be converted to half-metals in the antiferromagnetic state using an external transverse electric field. A transition from metal to semiconductor is observed by the edge passivation of the armchair nanoribbons. Zigzag $C_3N$ nanoribbons are magnetic metals when their edge atoms are not passivated by hydrogen. The magnetization is disappeared when the edge atoms are passivated. However, the kind of edge atoms controls the properties of the zigzag nanoribbons  so that the ribbons having just carbon atoms in one edge are metals. On the other hand, the zigzag ribbons with nitrogen atoms in both edges are semiconductors.
\section*{Conflicts of interest}
There are no conflicts to declare.

\section{Acknowledgment}
Authors are grateful to Dr. Hanif Hadipour for helpful suggestions.
\\

\bibliographystyle{model1a-num-names}
\bibliography{<your-bib-database>}

\end{document}